\documentclass{iau}
\usepackage{graphicx}
\usepackage{natbib}
\usepackage{graphicx}
\usepackage{float}
\usepackage{url}
\usepackage{subfigure}
\usepackage{wrapfig}
\usepackage{t1enc}

\newcommand{\hmpc}{\,$h^{-1}$\,Mpc}

\def\eg{{\frenchspacing\it e.g.}}

\def\be{\begin{equation}}
\def\ee{\end{equation}}
\def\ba{\begin{eqnarray}}
\def\ea{\end{eqnarray}}

\def\hmpc{h^{-1}\,{\rm Mpc}}

\title[Dynamics of the Cosmic Web] 
{Dynamics of pairwise motions in the Cosmic Web}
\author[Wojciech A. Hellwing]   
{Wojciech A. Hellwing$^{1,2}$}

\affiliation{$^1$Interdisciplinary Center for Mathematical and computational modelling (ICM),\\ University of Warsaw,  
ul. Pawi\'nskiego 2a, 02-186 Warsaw, Poland \\ email: {\tt wojciech.hellwing@durham.ac.uk} \\[\affilskip]
$^2$Institute for Computational Cosmology, University of Durham, \\ Science Site, South Road,
DH1-3LE Durham, UK}

\pubyear{2014}
\volume{308}  
\jname{The Zel'dovich Universe}
\editors{R. van de Weygaert, S. Shandarin, E. Saar, Jaan Einasto, eds.}
\begin{document}

\maketitle

\begin{abstract}
We present results of analysis of the dark matter (DM) pairwise velocity statistics in different
Cosmic Web environments. We use the DM velocity and density field from the Millennium 2 simulation
together with the NEXUS+ algorithm to segment the simulation volume into voxels uniquely identifying
one of the four possible environments: nodes, filaments, walls or cosmic voids. We show that the PDFs
of the mean infall velocities $v_{12}$ as well as its spatial dependence together with the perpendicular
and parallel velocity dispersions bear a significant signal of the large-scale structure 
environment in which DM particle pairs are embedded. The pairwise flows are notably colder 
and have smaller mean magnitude in wall and voids, when compared to much denser environments of filaments and nodes.
We discuss on our results, indicating that they are consistent with a simple theoretical predictions
for pairwise motions as induced by gravitational instability mechanism. Our results indicate that
the Cosmic Web elements are coherent dynamical entities rather than just temporal geometrical associations.
In addition it should be possible to observationally test various Cosmic Web finding algorithms by segmenting available
peculiar velocity data and studying resulting pairwise velocity statistics.

\end{abstract}

\firstsection 
\section{Introduction}
\label{sec:intro}

\begin{figure}[ht]
    \begin{center}
     \includegraphics[width=0.6\textwidth]{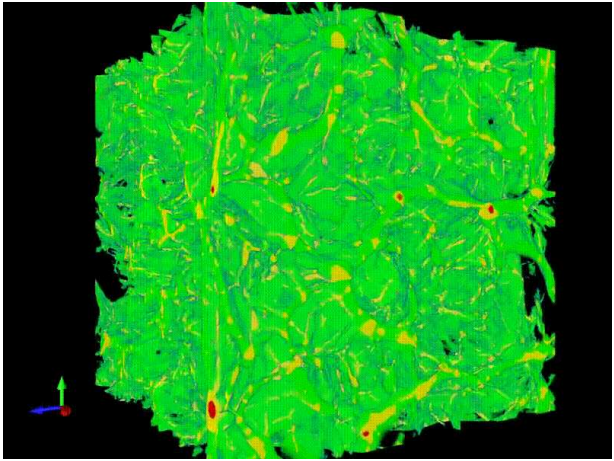}
    \end{center}  
  \caption{Rendering of the Cosmic Web elements from Millenium Simulation II as identified by the NEXUS algorithm.
         Different colours mark different elements of the Cosmic Web: dense nodes (clusters) are marked by red,
         elongated filaments are yellow, the pervading network of walls we depict by green and the remaining empty
         spaces correspond to cosmic voids.}
    	\label{fig:nexus_web}
\end{figure}

The Cosmic Web (CW) is the most salient manifestation of the anisotropic nature of gravitational collapse, the motor behind 
the formation of structure in the cosmos. Structure in the Universe has risen out of tiny primordial (Gaussian) density 
and velocity perturbations by means of gravitational instability. N-body computer simulations have profusely illustrated 
how the primordial density field transforms into a pronounced and intricate filigree of filamentary features, dented by 
dense compact clumps at the nodes of the network \citep{Bond1996}. Moreover, because of the hierarchical nature  
of the cosmic matter distribution, also filaments, sheets and voids emerge 
by the gradual merging of smaller scale specimen. In the process, small filaments align themselves along the direction of 
the emerging larger scale filament. The emerging picture is therefore one of a primordially and hierarchically defined skeleton 
whose weblike topology is imprinted over a wide spectrum of scales. Weblike patterns on ever larger scales get to dominate 
the density field as cosmic evolution proceeds, and as small scale structures merge into larger ones. On the other hand, within 
the gradually emptying void regions the topological outline of the early weblike patterns remains largely visible as a faint 
remnant of past glory (see {\it e.g.} Fig.~\ref{fig:nexus_web}).

So far most of the authors have focused on the clustering and density-based statistics analysis of the CW, as it is
seen in computer N-body simulations and galaxy catalogues 
\citep[\eg][]{Springel2006,AragonCalvo2010,Cautun_cweb_evo_2014,Metuki2014,Falck_cweb2014,Nuza_galaxy_cweb2014}. In this contribution
we want to adopt a different, yet complementary to previous studies, approach. Namely we will study the dynamics of the Cosmic
Web elements as reflected by the statistics of pairwise motions  observed inside different environments. The dynamical studies
of the CW elements are very important, both for providing complementary understanding their internal dynamics and kinematics,
and for established whether the observed large-scale structure environments are just reflection of the geometrical and spatial
galaxy distribution correlations or consists of a more dynamically coherent objects.

\section{The pairwise motions}
\label{sec:pairwise}

In this work we will consider lower-moments of the statistics of pairwise Dark Matter (DM) velocities as tracers
of the underlying dynamics driven by gravity in different CW environments.
The statistic of the mean relative pairwise velocity of galaxies $v_{12}$ -- 
the \textit{streaming velocity} -- reflects the ``mean tendency of well-separated galaxies to approach each other'' 
\citep{1980Peebles}. This statistic was
introduced by Davis\&Peebels \citep{DavisPeebles_BBGKY}  in the context of the kinetic BBGKY theory which describes the dynamical
evolution of a system of particles interacting via gravity.
In the fluid limit, its equivalent is the pair-density weighted relative velocity 
\be
\label{eqn:v12-wieghted}
\mathbf{v}_{12}(r) = \langle\mathbf{v}_1-\mathbf{v}_2\rangle_{\rho} = 
{\langle(\mathbf{v}_1-\mathbf{v}_2)(1+\delta_1)(1+\delta_2)\rangle\over 1+\xi(r)}\,\,,
\ee
where $\mathbf{v}_1$ and $\delta_1=\rho_1/\langle\rho\rangle-1$ stand for the peculiar velocity and fractional matter density 
contrast taken at 
point $\mathbf{r}_1$, $r=|\mathbf{r}_1-\mathbf{r}_2|$, and $\xi(r)=\langle\delta_1\delta_2\rangle$ is the 2-point density 
correlation function. 
The $\langle\cdots\rangle_{\rho}$ is the pair-weighted average, which differs from normal spatial averaging by the weighting factor 
$\mathcal{W}=\rho_1\rho_2/\langle\rho_1\rho_2\rangle$. Note $\mathcal{W}$ is proportional to the number density of pairs. The  
gravitational instability theory predicts that the $v_{12}(r)$ magnitude is governed by the 2-point correlation function $\xi(r)$ and 
the growth rate of matter density perturbations $f\equiv d\ln D_{+}/d\ln a$ (where $D_{+}(a)$ is the linear growing mode solution, 
and $a$ is the cosmological scale factor) through the pair conservation equation \citep{1980Peebles}. 
\cite{Juszkiewicz1999} provided a closed-form expression that is a good approximation to the solution of the pair conservation 
equation for universes with Gaussian initial conditions:
\be
v_{12}=-{2\over3}H_0rf\bar{\bar{\xi}}(r)[1+\alpha\bar{\bar{\xi}}(r)]\,\,,
\label{eqn:v12_ansatz}
\ee
where 
\be
\bar{\xi}(r)=(3/r^3)\int_0^r\xi(x)x^2dx\equiv\bar{\bar{\xi}}(r)[1+\xi(r)]\,\,.
\label{eqn:xi_spec}
\ee
Here, $\alpha$ is a parameter that depends on the logarithmic slope of $\xi(r)$ and $H_0=100\,h\,$km s$^{-1}$ Mpc$^{-1}$ is 
the present day value of 
the Hubble constant. It is clear that $v_{12}(r)$ is a strong function of $\xi(r)$ and $f$, which both in general will 
take different local averages in different CW environments. This motivates us to consider low-order moments of 
the pairwise velocity distribution
as tracers of CW specific local dynamics that should be reflected in motions of galaxies and DM. 
Specifically, in our analysis we will 
consider the following quantities:
\begin{itemize}
 \item the mean radial pairwise velocity, $v_{12}$; 
 \item the dispersion in the (radial) pairwise velocities (not centred), $\sigma_{\parallel}=\langle v_{12}^2\rangle^{0.5}$; 
 \item the mean transverse velocity of pairs, $v_{\perp}$; 
 \item the dispersion of the transverse velocity of pairs, $\sigma_{\perp}=\langle v_{\perp}^2\rangle^{0.5}$. 
\end{itemize}

It is well known that the cosmic velocity field has large coherence length, much 
larger than the density field \citep[\eg][]{Chod2002,Ciec2003}.
Thus the contribution of the large-scale velocity modes is always significant, when one concerns the peculiar velocities.
However this disadvantage (from the point of view of probing the velocity properties driven by local environment)
is not present for the pairwise velocity statistics we have just described above.
This is because at a given separation $r$ the velocity difference between a galaxy pair does not receive any net contribution from 
modes with wavelengths larger than the considered separation, since those give the same contribution for both particles/galaxies. 
Hence, at larger scales, for which the galaxies in a pair inhabit different haloes, the distribution of $v_{12}$ factorises into 
two individual peculiar velocity distributions for each galaxy/particle and those are always sensitive to non-linearities driven 
by virial motions within a given host halo alone (see \citet{Scoccimarro2004} for more details). Thanks to this,
the pairwise velocity statistics is automatically free of large-scale velocity modes contributions, and hence
is very well suited for the kind of analysis we want to perform.

\section{Dynamics of the Cosmic Web}
\label{sec:dynamics}

To study the dynamics of pairwise motions in different large-scale structure environments we use
the DM density and velocity fields from the Millennium-II simulation \citep{MS2}. To segment 
the simulation volume into voids, walls, filaments and cluster voxels we use the \verb#NEXUS+# algorithm \citep{Nexus}
applied on the DM density field computed on a $256^3$ grid (with a grid cell width of $0.39\hmpc$). For such defined
setup we obtain a catalogue of voxels fully segmenting the 3D simulation cube into four possible different CW
environments (see Fig.~\ref{fig:nexus_web}). The construction of 
the \verb#NEXUS+# algorithm is such that
the determination of the local environment is based on spatial variations along three Cartesian directions in the density field.
As a consequence
our voids correspond to spatial regions that are relatively empty with spherical-like symmetry, wall constitute thin planar 
structures, finally the filaments are elongated fibres. Our node environment corresponds to regions centred around massive
clusters and their local infall regions.

Having assessed the segmentation of our simulation volume into the CW elements we have computed all related pairwise 
statistics ($v_{12}$, $\sigma_{\parallel}$, $v_{\perp}$ and $\sigma_{\perp}$) for randomly Poisson sub-sampled DM particles.
We use $1/100$th of the original number of DM simulation particles. This sub-sampling leave us with a sufficient 
number of DM particles to study pairwise motions statistics for separations $\geq0.5\hmpc$, since at this separation we
have an averaged number of 10 particle pairs per voxel.
\begin{figure}[ht]
    \begin{center}
     \includegraphics[width=0.48\textwidth]{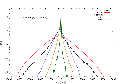}
     \includegraphics[width=0.48\textwidth]{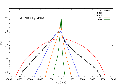}
    \end{center}  
  \caption{{\scriptsize The probability density functions for the mean pairwise streaming velocity $v_{12}$ computed at
  pair separation of $0.5\hmpc$ (left panel) and $2\hmpc$ (right panel). The line marking is the following:
  black with open circles - averaged for all cosmic environments ({\it i.e.} whole volume), red line - node/cluster
  environments, blue - filaments, orange - walls and we use green for cosmic voids.}}
    	\label{fig:pdf_v12}
\end{figure}

We begin by plotting the probability density functions for $v_{12}$ values computed at two different averaged pair
separations: $0.5$ and $2\hmpc$. The results are shown in the two panels of the Fig.~\ref{fig:pdf_v12}. The reference
point is marked always by a solid black line with open circle data points and corresponds to a result obtained from
the whole uniform volume ({\it i.e.} without segmenting into different morphology elements). The data corresponding
to the following environments we mark by: red (node/cluster), blue (filaments), orange (walls) and green for voids.
The both panels illustrate the striking difference of the $v_{12}$ distribution functions among different environments.
We denote that the width of the distribution (roughly corresponding here to dispersion) increase significantly
as we move from voids and walls towards denser environments of filaments and clusters. This is clearly visible at
both considered separations. This indicate that pairwise motions are much hotter in nodes and filaments than
in voids and walls. As we have described above, the mean streaming pairwise velocity is very sensitive 
to perturbation modes whose length is smaller than a considered pair separation. This property
of the pairwise statistics is now clearly consistent with our measured $v_{12}$ PDFs. It is evident that in environments 
like cosmic clusters and filaments the degree of non-linearity (in both the density and velocity field) is much
higher compared to relatively quieter cosmic walls and filaments. This difference is mostly driven by violent 
relaxation processes and thermalised virial motions that tend to dominate the peculiar velocity field inside
clusters and dense filaments regions \citep[\eg][]{Cautun_cweb_evo_2014,Nexus}. Secondly we can denote that the degree
of (a)symmetry of the PDFs are subject to a significant environment-driven variation. This can be especially seen
on the left panel of Fig.~\ref{fig:pdf_v12}, where we plot the probability density functions for pairs at $R_p=2\hmpc$
separation. The {\it stable clustering} model \citep{1980Peebles} predicts that at weakly non-linear scales
there should be slight excess of negative pairwise velocities reflecting the gravitational clustering mechanism that
still operates at those scales. Namely between two regimes of fully virialised motions at small scales and Hubble
expansion dominated at large scales there should be a region where some significant part of the peculiar motions 
still have potential character. We can observed this by noting that the cluster/node PDF function on the right panel
has much higher symmetry then the PDFs of the rest of the environments.  Such symmetric $v_{12}$ PDF is a characteristic 
of a fully randomised motions supported by velocity dispersion. This is the case inside virialised haloes, which clearly
dominate the pairwise flows signal in our node/cluster environments. 
\begin{figure}[ht]
    \begin{center}
     \includegraphics[width=0.48\textwidth]{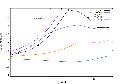}
     \includegraphics[width=0.48\textwidth]{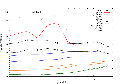}
    \end{center}  
  \caption{{\scriptsize The minus mean streaming velocities - $-v_{12}$ (left panel) and parallel 
  and perpendicular pair velocities dispersions  
  (right panel) - $\sigma_{\parallel}$, $\sigma_{\perp}$ - as a function of a pair separation $R_p$. 
  The lines colouring is consistent with the previous figure.}}
    	\label{fig:v12_sigmas_Rp}
\end{figure}

As a complementary probe of the pairwise motions we also study the minus mean infall velocities ($-v_{12}$) 
and parallel and perpendicular pair velocities dispersion ($\sigma_{\parallel}$, $\sigma_{\perp}$) as
a function of a pair separation $R_p$. We plot the corresponding results in the left and right panel
of the Fig.~\ref{fig:v12_sigmas_Rp}. We use the same colours as before to mark results for different
CW elements. On the left panel for comparison we also draw a purple line depicting the Hubble
expansion velocity. The results shown here are consistent with the previously discussed PDF properties.
We note both higher values of the mean infall velocities as well as much higher dispersions of both
the infall and perpendicular pair-velocity components for high density environments of nodes and filaments.
In contrast the pairwise motions in walls and especially in voids are characterised by very small
dispersions (are very cold). We can also denote another interesting observation regarding the
dynamics of pairwise motions in cosmic voids. Namely from the shape of the $-v_{12}(R_p)$ curve on the left
panel we can infer that, on average, the pairwise motions in voids have {\it super Hubble} character.
This means that inside the cosmic voids the expansion is faster then the background Hubble expansion rate.
Although this result was known for some time \citep[\eg][]{Seth2004}, as the inside regions of cosmic
voids can be treated locally as approximate Friedman Universes with lower effective $\Omega_m$ value
then the cosmic mean, it is very interesting to confirm this result in our data for pairwise motions.
The last thing we want the emphasise here is the lack of the offset between two different velocity directions
dispersions for the cosmic voids. This can be gauged by the offset between solid and dashed lines of the
same colour on the right panel of the Fig.~\ref{fig:v12_sigmas_Rp}. The offset between $\sigma_{\perp}$ and
$\sigma_{\parallel}$ is largest for the cluster environment and decrease as we move to filaments and walls.
However even for cosmic walls this offset is still significant (of the order of $\sim 33\%$), while 
in the case of the cosmic voids we observe that values and shapes of the both dispersions lines are 
fully consistent with each other. We speculate that this feature is driven by two facts: (i) the structure
formation has ceased inside cosmic voids, and (ii) there are no preferred directions for the voids segments
reflecting their average close-to spherical symmetry.

\section{Conclusions}

We have studied the dynamics and statistical properties of dark matter particles pairwise motions
in four different CW elements: nodes/clusters, filaments, walls and voids. We have shown that
both the probability density functions for the mean infall velocities as well as scale dependence 
of $v_{12}$ and two connected velocity dispersions are taking significantly different values and shapes
in different large-scale environments. The flows in voids and wall are much colder and also have
smaller magnitude, when compare to pairwise motions in denser node and filament environments. This indicate
that the \verb#NEXUS+# algorithm is very robust in determining different CW elements. The algorithm
does not use any velocity data whatsoever, and yet the CW elements delineated by this method
are characterised by pairwise motions whose statistics are consistent with simple theoretical predictions.

Our finding also indicate that the CW and its elements, as first defined by \citet{Bond1996}, is not only real
in the sense of the DM/galaxy density field morphology, but is also present at the level of peculiar motions
of DM. This is clearly seen once the contribution of large-scale velocity modes was removed, which can be
obtained for example by considering pairwise statistics as we did in this study. Such a strong correlations between
the distribution and dispersion of mean streaming velocities and the CW morphology also indicate
that the cosmic velocity-density relation has different limits of validity in different environments.
This reflects the different degree of non-linearity in the structure formation in voids, walls, filaments and nodes.

\section*{Acknowledgements}
This work would not be possible without a significant contribution of Marius Cautun, Rien van de Weygaert
and Carlos S. Frenk. I am very grateful for their help and many stimulating discussions.
The author acknowledge the support received from Polish National Science Center in grant 
no. DEC-2011/01/D/ST9/01960 and ERC Advanced Investigator grant of C.~S.~Frenk, COSMIWAY.
\bibliographystyle{hapj}
\bibliography{wah_bib}

\end{document}